\documentstyle[prl,aps,epsf]{revtex}
\twocolumn

\begin{document}
\draft
\title{Comment on ``Collective Excitations of a Bose-Einstein
Condensate \\ in a Magnetic  Trap''}
\maketitle

In a recent Letter, Mewes {\it et al } \cite{Mewes96Comm}
experimentally investigated
the collective excitations for  a Bose-Einstein condensate. For a nearly  
pure condensate they observed a damping time of $250(40)$ms for the
collective excitations at $30$Hz. They argued that for a nearly pure
condensate ($T\approx 0$) the
damping due to thermal contributions should be negligible and that so far 
there is no  theoretical  prediction for the damping of collective
excitations of a trapped condensate.    

In this Comment,  we shall calculate the damping of collective
excitations for their experiments by using the existing theories 
\cite{Hohenberg,Popov,Szepfalusy}. 
Let us consider a dilute gas model of $N$ weakly interacting bosons at finite
temperature with interaction  $v({\mathbf x}-{\mathbf x}^\prime )=
{4\pi a \hbar^2 \over m} 
\delta({\mathbf x}-{\mathbf x}^\prime)$ with $a$  the
$s$-wave scattering length. 
It is well known that the long-wavelength excitations ($k \rightarrow
0$) are phonons with the damped Bogoliubov spectrum 
$\omega= ck + i\gamma$ where the sound velocity $c \simeq {\hbar \over m}
\sqrt{4 \pi a n_0 }$.
The damping rate is given by Hohenberg and Martin \cite{Hohenberg} and
Popov \cite{Popov} for the low temperature and by Szepfalusy and Kondor
 \cite{Szepfalusy} for the intermediate temperature,
\begin{equation}
\label{eq:gamma}
\gamma = \left\{
	\begin{array}{ll}\displaystyle
   	{3\hbar k^5 \over 640\pi mn_0} + {3 \pi^3 (k_BT)^4 k \over
 	40 m n_0 c^4} & (T\ll T^*) \\ 
	\displaystyle
	{(k_BT)a k \over \hbar} \qquad 
	&(T^* \ll T \ll T_c)
	\end{array} \right. 
\end{equation}
where $T^* \equiv {4\pi a\hbar^2 n_0\over mk_B }$.
Next, we shall estimate the damping rate of
Ref.~\cite{Mewes96Comm}  using 
Eq.~(\ref{eq:gamma}).

First, we checked the temperature-independent part of the damping that
arises from the interaction between quasi-particles.  We 
found that for typical parameters (see below) 
$\gamma_0\equiv \gamma(T=0) 
 \sim 10^{-10}$s$^{-1}$ at an excitation frequency $\omega=2\pi \times 30$Hz,
which yields a decay time in the order 
of $10^{10}$s much longer than $250$ms found in
Ref.~\cite{Mewes96Comm}. Hereafter, $\gamma_0$ shall be ignored when
discussing the finite temperature case.

We plot  $\gamma$ against $T$ using 
(\ref{eq:gamma}) in Fig.~1, where the following numbers are used
\cite{Mewes96Comm,Mewes96b}: 
$a=65a_{Bhor}$, $n_0$ in the order of $10^{14}$cm$^{-3}$, the number
of atoms in 
condensate $N_0 = 5\times 10^6$, the  trap frequencies of
$250$Hz (radially) and $19$Hz (axially), and the sodium atom $m=23 \times 
1.66\times 10^{-27}$kg.
The temperature axis has been scaled against $T_c$ defined
by $T_c(N) \equiv {\hbar\bar{\omega} \over k_B}
(N/1.202)^{1/3} $ with $\bar{\omega} $ the
geometric mean of the harmonic trap frequencies
\cite{deGroot,Mewes96b}. Ketterle \cite{Ketterle} pointed out that 
a ``nearly pure condensate'' meant that the condensate fraction of
atoms was greater than or about $90\%$.
This infers that the total number of atoms $N \approx 5.5\times 10^6$,
leading to $T_c \simeq 0.84\mu$K.
The temperature can be implied  in the experiment 
 from the condensate fraction according to 
$N_0/N=1 -(T/T_c)^3$ \cite{deGroot} such that $T \simeq 0.5T_c$ 
\cite{Ketterle}.
With this temperature, we  find from Fig.~1 $\gamma\simeq 5.1$s$^{-1}$ and
$\gamma \simeq 3.6$s$^{-1}$ for $n_0=1.5\times 10^{14}$cm$^{-3}$ and
$n_0=3.0\times 10^{14}$cm$^{-3}$, respectively. 
In other words,   the
theoretical value of the decay time of collective excitations at
frequency $30$Hz  ranges from $190$ms to  $280$ms 
as $n_0$ is from  $1.5 \times 10^{14}$cm$^{-3}$ to $3.0
\times 10^{14}$cm$^{-3}$, consistent with
the experimental result $250(40)$ms.      

We therefore consider that the damping is caused by the interaction between
the collective excitation  and the thermal cloud rather than
the interaction between collective modes.
A more complete microscopic theory taking into account 
the inhomogeneity  and the presence of the harmonic
trap shall be published elsewhere.

The authors thank W. Ketterle for the very helpful
instruction on the experiments. 
W.V.L. is grateful to Prof. E. C. G. Sudarshan for his
criticisms and suggestions, and  thanks G. Ordonez for his meaningful
discussions. 

\begin{flushleft}
W. Vincent Liu and William C. Schieve\\
Physics Department and \\
~~Center for Statistical Mechanics \& Complex Systems \\
University of Texas \\
Austin, TX 78712\\
\end{flushleft}

\begin{figure}[htbp]
\begin{center}
\leavevmode
\epsfxsize=0.9\linewidth
\epsfbox{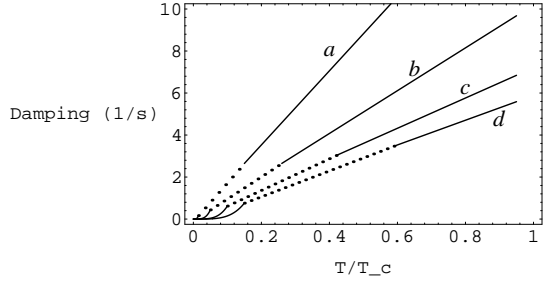}
\end{center}
\caption{The interpolation of the damping rate for different density of
the condensate. The
dotted lines mean that $T$ is near $T^*$ and 
simple analytical solutions are not yet available. 
$a.$ $n_0=0.5\times 10^{14}$cm$^{-3}$;  
$b.$ $n_0=1.5\times 10^{14}$cm$^{-3}$; 
$c.$ $n_0=3.0\times 10^{14}$cm$^{-3}$; 
 and $d.$ $n_0=4.5\times 10^{14}$cm$^{-3}$. } 
\end{figure}

\end{document}